\newcommand{\order}{\mathcal{O}}
\newcommand{\rel}{\mathrm{rel}}
\newcommand{\vegas}{\textsc{Vegas}}

\newcommand{\asqtad}{\textsc{ASQtad}}
\newcommand{\hisq}{\textsc{HISQ}}

\newcommand{\tree}{\mathrm{tree}}
\newcommand{\op}{Q}

\documentclass[color]{PoS}

\usepackage{amsmath}

\title{Radiative corrections to the m(oving)NRQCD action and heavy-light operators}

\ShortTitle{Radiative corrections in m(oving)NRQCD}

\author{%
  \mbox{\speaker{Eike H. M\"{u}ller}$^a$}, 
  \mbox{Christine T. H. Davies$^b$},
  \mbox{Alistair Hart$^a$},
  \mbox{Georg M. von Hippel$^c$},
  \mbox{Ron R. Horgan$^d$},
  \mbox{Iain Kendall$^b$},
  \mbox{Andrew Lee$^d$},
  \mbox{Stefan Meinel$^d$},
  \mbox{Chris Monahan$^d$},
  \mbox{Matthew Wingate$^d$}
  \vspace{2ex}\\
\llap{$^a$}
SUPA, School of Physics and Astronomy, University of Edinburgh, Edinburgh EH9 3JZ, UK\\
\llap{$^b$}
SUPA, Department of Physics and Astronomy, University of Glasgow, Glasgow G12 8QQ, UK\\
\llap{$^c$}
NIC, Deutsches Elektronen-Synchrotron DESY, Platanenallee 6,
15738 Zeuthen, Germany\\
\llap{$^d$}
DAMTP, University of Cambridge, Wilberforce Road, Cambridge CB3 0WA, UK
\vspace{4ex}\\
E-mail: \email{e.h.mueller@sms.ed.ac.uk}
}

\abstract{%
Rare decays of B mesons, such as $B \rightarrow K^*\gamma$ and $B\rightarrow K^{(*)}\ell^+\ell^-$ are loop suppressed in the Standard Model and sensitive to new physics. The final state meson in heavy-light decays at large recoil has sizeable momentum in the rest frame of the decaying meson. To reduce the resulting discretization errors we formulate the nonrelativistic heavy quark action in a moving frame. We discuss the perturbative renormalization of the leading order heavy-light operators in the resulting theory which is known as m(oving)NRQCD.

We also present radiative corrections to the NRQCD action computed using automated lattice perturbation theory. By combining this technique with high-beta simulations in the weak coupling regime of the theory higher order loop corrections can be calculated very efficiently.}

\FullConference{The XXVII International Symposium on Lattice Field Theory - LAT2009\\
		 July 26-31 2009\\
		 Peking University, Beijing, China}

\begin{document}

%%%%%%%%%%%%%%%%%%%%%%%%%%%%%%%%%%%%%%%%%%%%%%%%%%%%%%%%%%%%%%%%%%%
%
%       M A I N    T E X T
%
%%%%%%%%%%%%%%%%%%%%%%%%%%%%%%%%%%%%%%%%%%%%%%%%%%%%%%%%%%%%%%%%%%%

%%%%%%%%%%%%%%%%%%%%%%%%%%%%%%%%%%%%%%%%%%%%%%%%%%%%%%%%%%%%%%%%%%%
%%%%%%%%%%%%%%%%%%%%%%%%%%%%%%%%%%%%%%%%%%%%%%%%%%%%%%%%%%%%%%%%%%%
\section{Introduction}
%%%%%%%%%%%%%%%%%%%%%%%%%%%%%%%%%%%%%%%%%%%%%%%%%%%%%%%%%%%%%%%%%%%
%%%%%%%%%%%%%%%%%%%%%%%%%%%%%%%%%%%%%%%%%%%%%%%%%%%%%%%%%%%%%%%%%%%
Exclusive decays of $B$ mesons can be used to study the heavy flavour sector of the Standard Model. The magnitude of one of the least well known CKM matrix elements, $V_{ub}$, can be extracted from measurements of the decay $B\rightarrow \pi\ell\nu$. In addition, rare processes such as $B\rightarrow K^*\gamma$ and $B\rightarrow K^{(*)}\ell\ell$ provide an excellent opportunity to constrain new physics models. Recently, the experimental uncertainties in these processes have been reduced to below the $5\%$ level \cite{Nakao:2004th,Aubert:2008cy,Barberio:2008fa}.
Theoretical predictions of comparable precision are necessary to extract fundamental Standard Model parameters and to test new physics models.
%In \cite{Dalgic:2006dt} the vector form factors $f_+$ and $f_0$ have been calculated using an effective nonrelativistic heavy quark action and $|V_{ub}|$ is extracted from experimental results measured by the BaBar, Belle and CLEO experiments.

The calculation of hadronic heavy-light form factors involves several challenges: on currently available lattices, the Compton wavelength of the heavy quark is smaller than the lattice spacing and the relativistic quark action can not be used directly. Instead we work with a nonrelativistic effective heavy quark action; discretization errors are under control and can be reduced systematically by improving the action. If \mbox{$q=p_B-p_F$} is the momentum transfer between the initial and final state mesons then $q^2$ can be very small. For $B\rightarrow \pi\ell\nu$ most experimental data \cite{Athar:2003yg,Aubert:2006px,Hokuue:2006nr} comes from the large recoil region, and $q^2$ is zero for the radiative rare decay $B\rightarrow K^*\gamma$. This leads to substantial discretization errors in the final state meson due to its large momentum. These can be reduced by discretizing the nonrelativistic action in a moving frame of reference. The resulting formalism is known as m(oving)NRQCD \cite{Sloan:1997fc,Horgan:2009ti}.
%Rare decays pose additional challenges. The heavy-light transition is described by an effective continuum Hamiltonian which is obtained by integrating out physics at the electroweak scale. Four-quark operators generate nonlocal contributions which can not be treated in the framework of lattice QCD with currently available methods. Although it is argued that they are small \cite{Grinstein:2000pc,Wingate:2009PoS}, lattice calculations of the form factors of local operators need to be combined with other approaches such as QCD sum rules.

The heavy quark action used in this work is correct up to $\order(1/m^2,v_{\rel}^4)$, where $v_{\rel}$ is the relative velocity between the two quarks in a heavy-heavy meson. One source of systematic errors is the mismatch between the ultraviolet modes in the continuum and lattice mNRQCD. In heavy-heavy spectrum calculations radiative corrections are of comparable size to relativistic corrections \cite{Gray:2005ur}. Typical loop momenta are of the order of the heavy quark mass, where the strong coupling constant is small. In this work we calculate radiative corrections to the mNRQCD action and heavy-light operators. The calculations are carried out at one loop in mean field improved diagrammatic perturbation theory. Due to the complexity of the action the Feynman rules are generated using an automated expansion algorithm \cite{Luscher:1985wf,Hart:2004bd,Hart:2009nr}
and phase space integrals are solved numerically with the adaptive Monte Carlo integrator \vegas\ \cite{Lepage:1977sw}.

In addition, fundamental parameters of the Standard Model, such as the heavy quark mass in the $\overline{MS}$ scheme, $m_b^{\overline{MS}}$, can be extracted from lattice calculations if their perturbative expansion is known to high orders both in the continuum and on the lattice.
In this work we use a mixed strategy for calculating higher order loop corrections to the heavy quark self-energy: by simulating the quenched theory in the weak coupling regime (corresponding to large values of the inverse coupling $\beta$) and fitting the results to an expansion in $\alpha_s$, the gluonic higher order loop corrections can be obtained \cite{Dimm:1994fy}. This has to be supplemented with the calculation of the fermionic contribution in diagrammatic perturbation theory. At one loop order the fermionic vacuum polarisation does not contribute and we compare results from diagrammatic perturbation theory and high-$\beta$ simulations.
%As there is no fermionic contribution to the vacuum polarisation at one loop, this calculation can be used to compare high-$\beta$ results to those from diagrammatic perturbation theory. In addition, the high-$\beta$ fits can be improved by constraining the one loop coefficient to the value found in digrammatic calculations.

%%%%%%%%%%%%%%%%%%%%%%%%%%%%%%%%%%%%%%%%%%%%%%%%%%%%%%%%%%%%%%%%%%%
\subsection{Effective theories for heavy quarks on the lattice}
%%%%%%%%%%%%%%%%%%%%%%%%%%%%%%%%%%%%%%%%%%%%%%%%%%%%%%%%%%%%%%%%%%%
After integrating out fluctuations at the $b$ quark scale, the Lagrangian of the nonrelativistic heavy quark theory can be written as an expansion in the inverse heavy quark mass. A mNRQCD action which is correct to $\order(1/m^2,v_{\rel}^4)$ is derived in detail in \cite{Horgan:2009ti}. On the lattice it can be written as

\begin{equation}
  S = \sum_{\vec{x},\tau}\psi^+(\vec{x},\tau)\Big[
  \psi(\vec{x},\tau) - \left(
    1-\frac{\delta H}{2}
  \right)
  \left(
  	1-\frac{H_0}{2n}
  \right)^n
  U_4^\dagger
    \left(
  	1-\frac{H_0}{2n}
  \right)^n
  \left(
    1-\frac{\delta H}{2}
  \right)\psi(\vec{x},\tau-1)\Big].
\end{equation}
%\begin{equation}
%  \begin{split}
%  S &= \sum_{\vec{x},\tau}\psi^+(\vec{x},\tau)\Big[
%  \psi(\vec{x},\tau) - \left(
%    1-\frac{\delta H}{2}
%  \right)
%  \left(
%  	1-\frac{H_0}{2n}
%  \right)^n
%\label{sec:HQ_lattice:NRQCDkernel}
%\\&\quad\times\;\;
%  U_4^\dagger(\vec{x},\tau-1)
%  \left(
%  	1-\frac{H_0}{2n}
%  \right)^n
%  \left(
%    1-\frac{\delta H}{2}
%  \right)\psi(\vec{x},\tau-1)\Big].
%  \end{split}
%\end{equation}
The lowest order kinetic term is
$H_0 = - i \vec{v}\cdot\vec{\Delta}^\pm
    -\left(\Delta^{(2)}-\Delta_v^{(2)}\right)/(2\gamma m)$ 
%\begin{equation}
%H_0 = - i \vec{v}\cdot\vec{\Delta}^\pm
%    -\frac{\Delta^{(2)}-\Delta_v^{(2)}}{2\gamma m} \label{eqn:HQ_lattice:mNRQCDH0}
%\end{equation}
where $\vec{v}$ is the frame velocity and 
$\Delta_j^\pm$, $\Delta^{(2)}$ and $\Delta_v^{(2)}$ are first and second order covariant finite difference operators.
$\delta H$ contains higher order corrections in $1/m$ and operators which reduce discretization artifacts.
%%%%%%%%%%%%%%%%%%%%%%%%%%%%%%%%%%%%%%%%%%%%%%%%%%%%%%%%%%%%%%%%%%%
%%%%%%%%%%%%%%%%%%%%%%%%%%%%%%%%%%%%%%%%%%%%%%%%%%%%%%%%%%%%%%%%%%%
\section{Perturbative renormalization of m(oving)NRQCD}
\begin{figure}
 \begin{center}
   \includegraphics[width=0.5\linewidth]{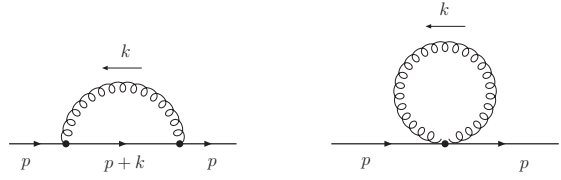}
   \caption{One loop corrections to the heavy quark self-energy.}
   \label{fig:selfenergy_diagrams}
 \end{center}
\end{figure}

%%%%%%%%%%%%%%%%%%%%%%%%%%%%%%%%%%%%%%%%%%%%%%%%%%%%%%%%%%%%%%%%%%%
%%%%%%%%%%%%%%%%%%%%%%%%%%%%%%%%%%%%%%%%%%%%%%%%%%%%%%%%%%%%%%%%%%%
Renormalization constants of kinetic terms in the mNRQCD action can be obtained from the heavy quark self-energy, which at one loop is given by the two diagrams in Fig. \ref{fig:selfenergy_diagrams}. We include mean field corrections by dividing each link in the action by its mean value \mbox{$u_0=1-\alpha_s u_0^{(2)} + \dots$}.
%This reduces the size of radiative corrections significantly.
%Feynman rules are automatically generated using the \hippy\ code \cite{Hart:2004bd} and the resulting phase space integrals are solved by the adaptive Monte Carlo integrator \vegas. 
%Calculating derivatives of a specific diagram with respect to external momenta is straightforward as the Leibniz rule is overloaded for the multiplication of \taylur\ \cite{vonHippel:2005dh} objects.
%%%%%%%%%%%%%%%%%%%%%%%%%%%%%%%%%%%%%%%%%%%%%%%%%%%%%%%%%%%%%%%%%%%
\subsection{Renormalization parameters}
\begin{figure}
\begin{center}
  \includegraphics[angle=270,width=0.7\linewidth]{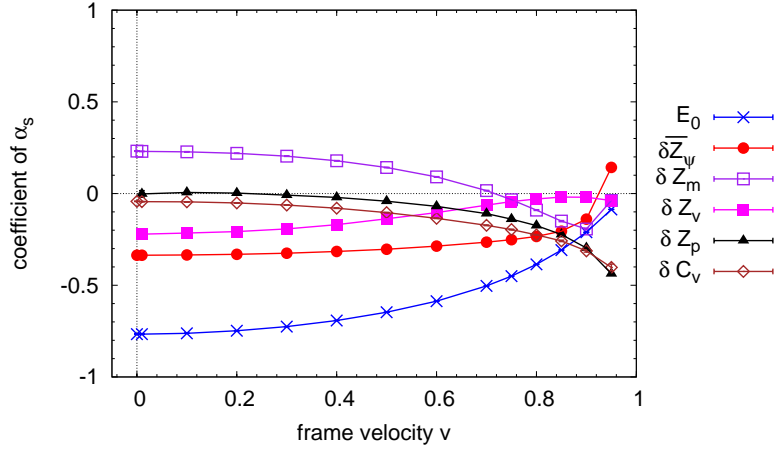}
  \caption{One loop corrections to heavy quark renormalization parameters. Mean field corrections are included. The heavy quark mass is $m=2.8$ and the stability parameter $n=2$.}
  \label{fig:ren_parm}
\end{center}  
\end{figure}

%%%%%%%%%%%%%%%%%%%%%%%%%%%%%%%%%%%%%%%%%%%%%%%%%%%%%%%%%%%%%%%%%%%
We calculate the one loop corrections to the zero point energy $E_0$, wavefunction renormalization\footnote{We only show the infrared finite part without the logarithmic divergence $-2/(3\pi)\log a^2\lambda^2$.} $\delta \overline{Z}_\psi$, heavy quark mass $\delta Z_m$, frame velocity $\delta Z_v$ and the energy shift between lattice mNRQCD and continuum QCD $\delta C_v$ for a range of frame velocities, see Fig. \ref{fig:ren_parm}. The calculation is performed in an infinite volume with a small gluon mass $\lambda$ as an infrared regulator. All renormalization parameters are obtained from the self-energy by taking appropriate derivatives with respect to the external quark momentum \cite{vonHippel:2005dh,vonHippel:2007xd}. Of particular interest is the renormalization of the external momentum $\delta Z_p$, which in the continuum is protected by reparametrization invariance. On the lattice, it is small for not too large frame velocities. The gluon action is Symanzik improved in all calculations presented here.
After including mean field corrections, the magnitude of the one loop corrections is of order one and smaller, as expected for a well behaved perturbative expansion.

%%%%%%%%%%%%%%%%%%%%%%%%%%%%%%%%%%%%%%%%%%%%%%%%%%%%%%%%%%%%%%%%%%%
\subsection{High-beta simulations}
%%%%%%%%%%%%%%%%%%%%%%%%%%%%%%%%%%%%%%%%%%%%%%%%%%%%%%%%%%%%%%%%%%%
Calculation of higher order loop corrections becomes increasingly difficult due to the large number of diagrams and the complicated structure of vertices with a large number of gluons. Instead we can calculate radiative corrections by measuring the heavy quark two point function at large values of $\beta$, extract the renormalization parameters and fit them to a polynomial in $\alpha_s$ \cite{Dimm:1994fy}. Nonperturbative contributions are suppressed by using twisted boundary conditions \cite{Luscher:1985wf,'tHooft:1979uj}. As the inclusion of fermionic vacuum polarisation effects makes the generation of configurations computationally expensive, the configurations used in our calculation are quenched. The missing fermionic corrections can be included relatively easily as they require the evaluation of a small number of diagrams.
In addition, the quality of the polynomial fit can be improved if the one loop coefficient is constrained to the value calculated in diagrammatic perturbation theory.

In Tab. \ref{tab:ren_parm_finitevolumeE0} we show the results of a polynomial fit in $\alpha_s$. In the first case, the one loop coefficient was unconstrained, whereas in the second case it was constrained to the value from diagrammatic perturbation theory. For all but the highest frame velocity the  one loop coefficients agree within statistical errors.

\begin{table}
 \begin{center}
%%%%%%%% Table automatically generated on Fri, 03 Jul 2009 13:06:16 +0000
%%%%%%%% renormalisation parameter: deltaE0
\begin{tabular}{ccp{2.0cm}p{2.0cm}p{2.0cm}p{2.0cm}}
\hline
 fit & & $v=0.0$  & $v=0.2$  & $v=0.4$  & $v=0.8$ \\ \hline\hline
unconstrained & 1 loop
  & $-2.6321(69)$ & $-2.5797(67)$ & $-2.4146(64)$ & $-1.812(16)$ \\[0ex]
  & 2 loops
  & $-0.24(36)$ & $-0.33(36)$ & $-0.93(34)$ & $-0.01(67)$ \\
  & 3 loops
  & $-15.1\pm3.3$ & $-14.8\pm3.3$ & $-8.9\pm3.3$ & $-20.0\pm5.2$
  \\\hline
constrained
  & 1 loop
  & $-2.6254$ & $-2.5732$ & $-2.4140$ & $-1.7284$ \\
  & 2 loops
  & $-0.55(17)$ & $-0.63(17)$ & $-0.96(15)$ & $3.06(59)$ \\
  & 3 loops
  & $-12.7\pm 2.1$ & $-12.4\pm 2.3$ & $-8.6\pm 2.1$ & $0.3\pm 5.9$
  \\\hline
\end{tabular}
%%%%%%%% Table automatically generated on Fri, 03 Jul 2009 13:06:16 +0000
  \caption{Coefficients of the perturbative expansion of the zero point energy shift $E_0$ for a simple mNRQCD action with $\delta H=0$. The calculation was carried out on a $18\times 6^3$ lattice and results are shown both for unconstrained and constrained one loop coefficients. Mean field corrections are not included. The heavy quark mass is $m=2.0$ and the stability parameter $n=2$. All results are preliminary.}
  \label{tab:ren_parm_finitevolumeE0}
 \end{center}
\end{table}

%%%%%%%%%%%%%%%%%%%%%%%%%%%%%%%%%%%%%%%%%%%%%%%%%%%%%%%%%%%%%%%%%%%
\subsection{Higher order kinetic terms}
%%%%%%%%%%%%%%%%%%%%%%%%%%%%%%%%%%%%%%%%%%%%%%%%%%%%%%%%%%%%%%%%%%%
We also investigate the breaking of rotational invariance by higher order kinetic terms in the NRQCD Lagrangian. Energy splittings between mesons with the same $\vec{p}^2$ but different $\sum_j p_j^4$ have been calculated nonperturbatively in NRQCD.
Although these splittings are small compared to the
total kinetic energy, they are not compatible with zero within errors. They can be removed by tuning the coefficient of the rotationally noninvariant term $\Delta^{(4)}/(24m)$ to $c_5=2.6$, which has to be compared to the tree level value of $c_5^{\tree}=1$.

\begin{figure}
\begin{center}
  \includegraphics[angle=270,width=0.6\linewidth]{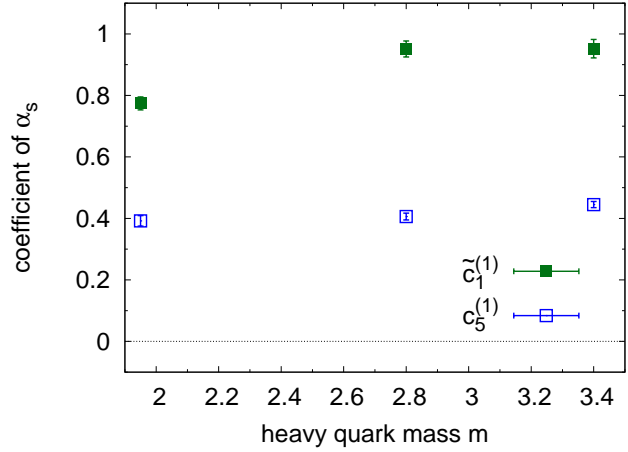}
  \caption{One loop renormalization parameters of higher order kinetic terms in the NRQCD action. Mean field corrections are included. The stability parameter is $n=4$.
  %Errors are statistical from the \vegas\ integration.
}
  \label{fig:higher_order_kinetic}
\end{center}  
\end{figure}

In Fig. \ref{fig:higher_order_kinetic} the one loop corrections in the perturbative calculation are shown, together with the corresponding corrections to the coefficient $\tilde{c}_1$ which multiplies the operator $-\left(\Delta^{(2)}\right)^2/(8m^3)$.
As can be seen from this plot, the smallness of the one loop corrections is not compatible with the nonperturbative calculation. Effects of higher order derivative operators in the action and the size of discretization errors in the nonperturbative simulations are currently being explored.
%%%%%%%%%%%%%%%%%%%%%%%%%%%%%%%%%%%%%%%%%%%%%%%%%%%%%%%%%%%%%%%%%%%
%%%%%%%%%%%%%%%%%%%%%%%%%%%%%%%%%%%%%%%%%%%%%%%%%%%%%%%%%%%%%%%%%%%
\section{Radiative corrections to heavy-light currents}
%%%%%%%%%%%%%%%%%%%%%%%%%%%%%%%%%%%%%%%%%%%%%%%%%%%%%%%%%%%%%%%%%%%
%%%%%%%%%%%%%%%%%%%%%%%%%%%%%%%%%%%%%%%%%%%%%%%%%%%%%%%%%%%%%%%%%%%
We calculate the one loop correction to the vector $(V)$ and tensor $(T)$ current. 
\begin{equation}
  \op^{(V)\mu} = \overline{q} \gamma^\mu\Psi,\qquad
  \op^{(T)\mu\nu} = \frac{e}{16\pi^2}m\overline{q} \sigma^{\mu\nu}\Psi.
\end{equation}
The branching ratio of the decay $B\rightarrow\pi\ell\nu$ can be calculated if the form factor of the vector current is known. The local contributions to the rare decays $B\rightarrow K^*\gamma$ and $B\rightarrow K^{(*)}\ell\ell$ are given by the tensor and vector current respectively.

The QCD field $\Psi$ has to be expressed in terms of fields in the effective theory in the moving frame. At leading order in the $1/m$ expansion two operators $\op_\pm^{(\Gamma)}$ contribute on the lattice for each Dirac structure $\Gamma$. From these a lattice operator $\op^{(\Gamma)}_{\mathrm{lat}}$ can be constructed as
\begin{equation}
\op^{(\Gamma)}_{\mathrm{lat}} = (1+\alpha_s c_+^{(\Gamma)})
    \op_+^{(\Gamma)} + \alpha_s c_-^{(\Gamma)} \op_-^{(\Gamma)}.
\end{equation}
The matching coefficients $c_{\pm}^{(\Gamma)}$ are adjusted such that matrix elements in lattice mNRQCD and continuum QCD agree at one loop, 
$\langle s |\op^{(\Gamma)}| b\rangle_{\mathrm{QCD}}
    = \langle s |\op^{(\Gamma)}_{\mathrm{lat}}| b\rangle_{\mathrm{lat}} + \order(\alpha_s^2,1/m)$. This implies
\begin{equation}
% \begin{split}
    c_+^{(\Gamma)}(am,\mu/m) = \delta Z^{(\Gamma)}(\mu/m) - \delta \tilde{Z}^{(\Gamma)}_{++}(am),\qquad\qquad    c_-^{(\Gamma)}(am) = - \delta \tilde{Z}^{(\Gamma)}_{+-}(am).
%\end{split}
\end{equation}
$\delta Z^{(\Gamma)}(\mu/m)$ is the one loop correction to the continuum operator $\op^{(\Gamma)}$ whereas the one loop mixing matrix between the operators $\op^{(\Gamma)}_\pm$ on the lattice is $\delta\tilde{Z}^{(\Gamma)}(am)$.

\begin{figure}
 \begin{center}
% ...
  \begin{minipage}{0.49\linewidth}
    \begin{center}
      \includegraphics[angle=270,width=\linewidth]{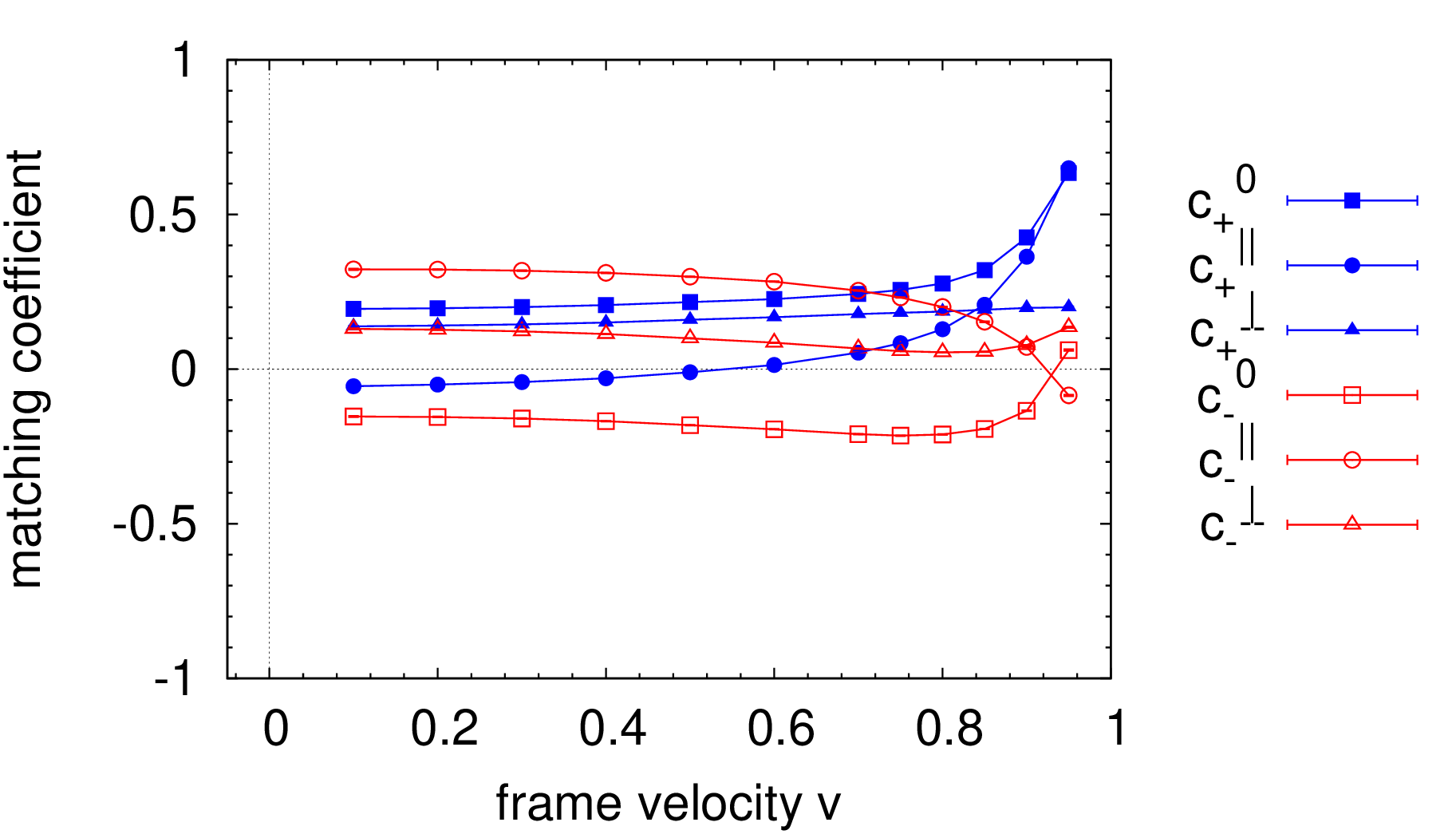}
    \end{center}
  \end{minipage} \hfill
% ...
  \begin{minipage}{0.49\linewidth}
    \begin{center}
      \includegraphics[angle=270,width=\linewidth]{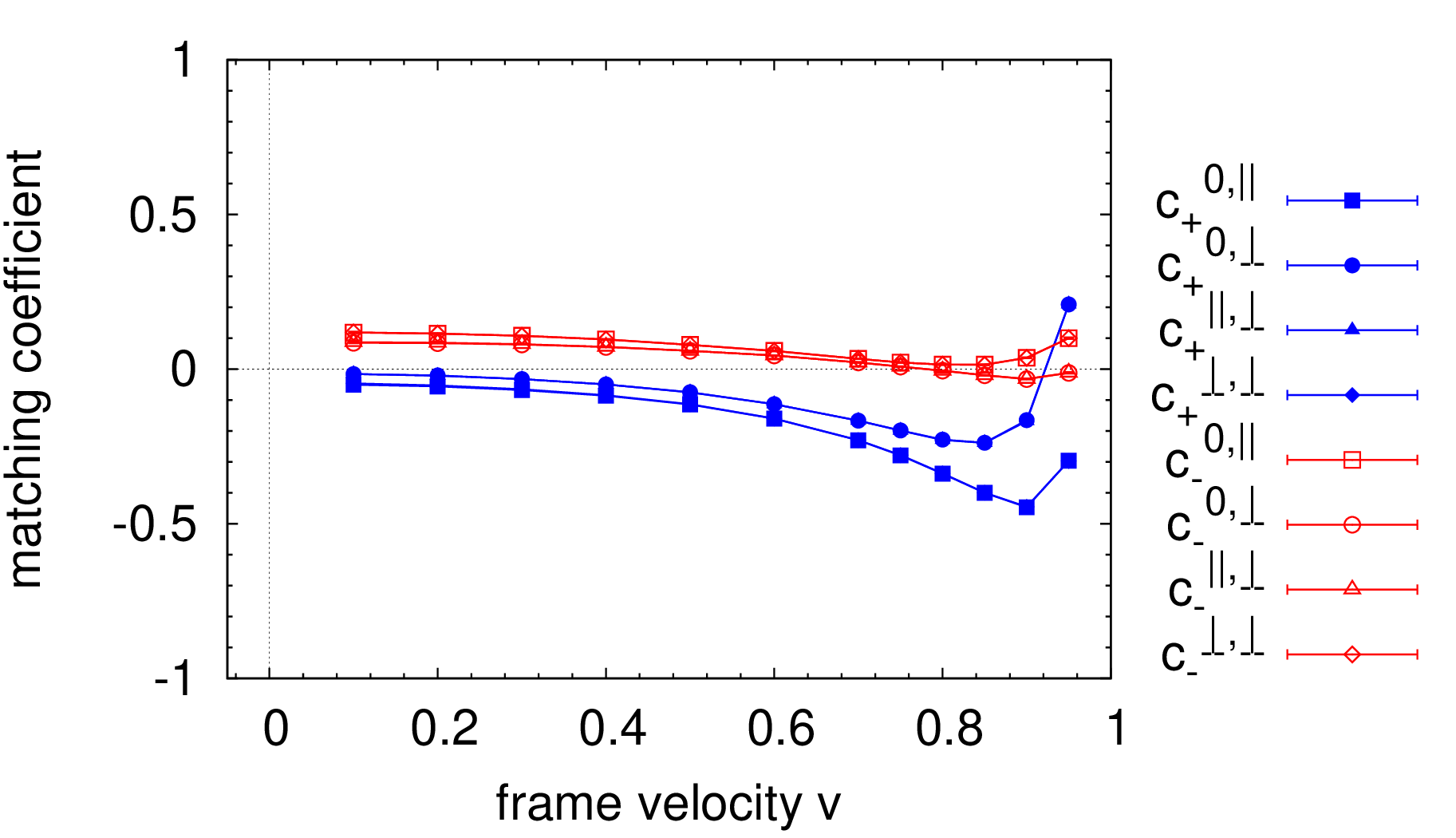}
    \end{center}
  \end{minipage}
% ...
  \caption{One loop matching coefficients for the vector (left) and tensor (right) current. The Lorentz indices can be timelike ($0$), parallel ($\parallel$) or perpendicular ($\perp$) to the frame velocity. Mean field corrections are included, the massless light quark is discretized in the \asqtad\ action. The heavy quark mass is $m=2.8$ and the stability parameter $n=2$; the renormalization scale for the tensor current is $\mu=m$.}
  \label{fig:matchingcoefficients}
 \end{center}
\end{figure}

Numerical results for the matching coefficients of the vector and tensor current are shown in Fig. \ref{fig:matchingcoefficients}. The light quark is discretized using the \asqtad\ action and its mass is set to zero in the matching calculation.
Nonperturbative form factors of these currents have been calculated for a simple mNRQCD action \cite{Meinel:2008th}. This calculation is currently repeated with the full $\order(1/m^2,v_{\rel}^4)$ mNRQCD action used in this work \cite{Liu:2009PoS}, including $1/m$ corrections to the currents at tree level. 
%%%%%%%%%%%%%%%%%%%%%%%%%%%%%%%%%%%%%%%%%%%%%%%%%%%%%%%%%%%%%%%%%%%
%%%%%%%%%%%%%%%%%%%%%%%%%%%%%%%%%%%%%%%%%%%%%%%%%%%%%%%%%%%%%%%%%%%
\section{Conclusion and Outlook}
%%%%%%%%%%%%%%%%%%%%%%%%%%%%%%%%%%%%%%%%%%%%%%%%%%%%%%%%%%%%%%%%%%%
%%%%%%%%%%%%%%%%%%%%%%%%%%%%%%%%%%%%%%%%%%%%%%%%%%%%%%%%%%%%%%%%%%%
%The HPQCD collaboration uses an effective nonrelativistic action to discretise heavy quarks on the lattice. The theory is formulated in a moving frame of reference to reduce the discretisation errors in the hadronic final state of heavy-light decays at high recoil such as $B\rightarrow\pi\ell\nu$ or the rare decays $B\rightarrow K^*\gamma$ and $B\rightarrow K^{(*)}\ell\ell$. Systematic errors in this approach need to be reduced to achieve sufficient precision in the calculation of energy levels and hadronic form factors. 

We calculated radiative corrections to leading order kinetic terms in the mNRQCD action on the lattice in diagrammatic perturbation theory both in an infinite volume and on a finite lattice with twisted boundary conditions. The one loop coefficients are compared to those from high-$\beta$ simulations and used to stabilise the polynomial fit in the strong coupling constant. We find good agreement between the two methods for a reasonable range of frame velocities.

Radiative corrections to rotationally noninvariant higher order kinetic terms are calculated in NRQCD. Further work is necessary to reduce the mismatch of one loop results and nonperturbative calculations of the matching coefficients. Results for the one loop corrections to the leading order operators of the heavy-light vector and tensor currents in mNRQCD are presented. These will be combined with the nonperturbative calculation of the form factors which is currently being carried out with the full $\order(1/m^2,v_{\rel}^4)$ action used in perturbative calculations.

A two loop calculation of the fermionic contribution to the heavy quark self-energy is currently being carried out and will complement the gluonic contributions from quenched high-$\beta$ simulations. Other terms in the action, most importantly the $\vec{\sigma}\cdot\vec{B}/(2m)$ chromomagnetic term which affects the hyperfine splitting, need to be renormalized. Our approach for evaluating Feynman diagrams is flexible enough to extend it to other actions; so far we have repeated the current matching calculation with the \hisq\ light quark action but other combinations such as static heavy quarks decaying to domain wall fermions could be investigated as well.
%%%%%%%%%%%%%%%%%%%%%%%%%%%%%%%%%%%%%%%%%%%%%%%%%%%%%%%%%%%%%%%%%%%
%%%%%%%%%%%%%%%%%%%%%%%%%%%%%%%%%%%%%%%%%%%%%%%%%%%%%%%%%%%%%%%%%%%
\acknowledgments
%%%%%%%%%%%%%%%%%%%%%%%%%%%%%%%%%%%%%%%%%%%%%%%%%%%%%%%%%%%%%%%%%%%
%%%%%%%%%%%%%%%%%%%%%%%%%%%%%%%%%%%%%%%%%%%%%%%%%%%%%%%%%%%%%%%%%%%
We would like to thank Lew Khomskii, Zhaofeng Liu and Junko Shigemitsu for useful discussions.
This work has made use of the resources provided by the Edinburgh Compute and Data Facility (ECDF) and the Cambridge High Performance Computing Service. The ECDF is partially supported by the eDIKT initiative.
We also acknowledge support from the DEISA Extreme Computing Initiative (DECI). 
%%%%%%%%%%%%%%%%%%%%%%%%%%%%%%%%%%%%%%%%%%%%%%%%%%%%%%%%%%%%%%%%%%%
%
%       R E F E R E N C E S
%
%%%%%%%%%%%%%%%%%%%%%%%%%%%%%%%%%%%%%%%%%%%%%%%%%%%%%%%%%%%%%%%%%%%

\end{document}